\documentclass[prd,aps,showkeys,two column]{revtex4}

\usepackage{amsmath}
\usepackage{amssymb}
\usepackage{amsthm}
\usepackage{mathrsfs}
\usepackage{graphicx}
\usepackage{fancyhdr}
\usepackage{array}
\usepackage{hyperref}
\usepackage[all]{xy}
\usepackage{eufrak}
\usepackage{euscript}
\usepackage{enumerate}
\usepackage{dsfont}

\newcommand{\be}{\begin{equation}}
\newcommand{\ee}{\end{equation}}
\newcommand{\bs}{\begin{split}}
\newcommand{\es}{\end{split}}

\def \v {\vec}

\begin{document}


\title{Dual approaches for defects condensation}

\author{$^{a,c}$L. S. Grigorio, $^{b}$M. S. Guimaraes, $^{a}$R. Rougemont, $^{a}$C. Wotzasek}

\affiliation{
$^{a}$Instituto de F\'\i sica, Universidade Federal do Rio de Janeiro,
\\21941-972, Rio de Janeiro, Brazil
\\
$^{b}$Departamento de F\'\i sica Te\'orica, Instituto de F\'\i sica, UERJ - Universidade do
Estado do Rio de Janeiro, Rua S\~ ao Francisco Xavier 524, 20550-013 Maracan\~ a, Rio de Janeiro,
Brazil
\\
$^{c}$Centro Federal de Educa\c{c}\~ao Tecnol\'ogica Celso Suckow da Fonseca
\\28635-000, Nova Friburgo, Brazil
}

%
%
%

\date{\today}

\begin{abstract}
We review two methods used to approach the condensation of defects phenomenon. Analyzing in details
their structure, we show that in the limit where the defects proliferate until occupy the whole
space these two methods are dual equivalent prescriptions to obtain an effective theory for the
phase where the defects (like monopoles or vortices) are completely condensed, starting from the
fundamental theory defined in the normal phase where the defects are diluted.
\end{abstract}




\maketitle


\section{Introduction}
The quantum field theory description of a physical system relies on a proper identification of its
degrees of freedom which are then interpreted as excited states of the fields defining the theory.
However it is sometimes the case that the theory may contain important structures which are not
described in this way and cannot be expressed in a simple manner in terms of the fields appearing
in the Lagrangian, having a non-local expression in terms of them. These structures appear under
certain conditions as {\it defects}; prescribed singularities of the fields defining the theory. A
general conjecture \cite{montol} claims that defects are described by a dual formulation in which
they appear as excitations of the dual field, but this can be proved only in some particular
instances. Nevertheless, much can be gained just with the information that these structures appear
as singularities of the fundamental fields even without knowing their precise dynamics. A pressing
question is if it is possible to address, with this limited information, the situation in which
the collective behavior of defects becomes the dominant feature of a theory. It is one of the
purposes of this work to discuss an extreme case of sorts. We want to present a general proposal
of how to describe a situation in which the singularities of the fields proliferate defining a new
vacuum for the system. In this picture the new degrees of freedom are recognized as excitations of
the established condensate of defects.

This view is supported by the fact that if we are interested only in the low lying excitations it
is perfectly reasonable to take the condensate as given, not worrying how it was set on, and
construct an effective field theory describing the excitations. It is well known for instance that
the pions, which can be recognized as excitations of the chiral symmetry breaking condensate
composed of quark-antiquark pairs, can be described by an effective field theory without knowing
about QCD. Even though we need not know the details of how the condensate is formed it is
important to stress that the condensate defines the vacuum and carries vital information about the
symmetry content used in the construction of the effective theory. It is in this way also bound
to have an effect in all the other fields comprising the system. The example of a superconducting
medium also comes to mind, where the condensate vacuum endows the electromagnetic excitations with
a mass. This same idea is employed on the electroweak theory where a condensate is the only
consistent way to give mass to the force carriers, the $W$ and $Z$, and in fact to account for all
the masses of the standard model. This is an example where the properties of the condensate itself
are not completely established and still a matter of debate. The currently accepted view is that
its low lying excitations are the Higgs particles, still to be detected, described by a scalar
field.

More akin to our take on the condensate concept, as a collective behavior of defects, is the dual
superconductor model of confinement which is based on the superconductor phenomenology \cite{intro}.
It is expected that the QCD vacuum at low energies is a chromomagnetic condensate leading to the
confinement of color charges immersed in this medium. In dual superconductor models of color
confinement, magnetic monopoles appear as topological defects in points of the space where the
abelian projection becomes singular \cite{ripka}. There are in fact many other examples in which
the condensation of defects is responsible for drastic changes in the system by defining the new
vacuum of the theory. We may mention vortices in superfluids and line-like defects in solids which
are responsible for a great variety of phase transitions \cite{mvf}. All these instances point
to the importance of getting a better understanding of the condensation phenomenon.

In all these examples there are some general features of the condensates which can tell a lot about
what to expect of the system when condensation sets in without the precise knowledge of how this
happened. These general features are what we intend to explore in this Letter. The main inspiration
for this work comes from the study of two particular approaches to this problem: one is the Abelian
Lattice Based Approach (ALBA) discussed by Banks, Myerson and Kogut in \cite{Banks:1977cc}
within the context of relativistic lattice field theories and latter also by Kleinert in
\cite{tric1} in the condensed matter context. The other one is the Julia-Toulouse Approach (JTA)
introduced by Julia and Toulouse in \cite{j-t} within the context of ordered solid-state media and
later reformulated by Quevedo and Trugenberger in the relativistic field theory context \cite{QT}.

The ALBA was used, for example, by Banks, Myerson and Kogut to study phase transitions in abelian
lattice gauge theories \cite{Banks:1977cc}. A few years latter Kleinert obtained a disorder field
theory for the superconductor from which he established the existence of a tricritical point
separating the first-order from the second-order superconducting phase transitions \cite{tric1}.
In this Letter we shall be using the notations in the recent book by Kleinert \cite{mvf}.

Developing in the work of Julia and Toulouse, Quevedo and Trugenberger studied the different
phases of field theories of compact antisymmetric tensors of rank $h-1$ in arbitrary
space-time dimensions $D=d+1$. Starting in a coulombic phase, topological defects of dimension
$d-h-1$ ($(d-h-1)$-branes) may condense leading to a confining phase. In that work one of the
applications of the JTA was the explanation of the axion mass.  It was known that the QCD instantons
generate a potential which gives mass to the axion. However, the origin of this mass in a dual
description were a puzzle. When the JTA is applied it is clear that the condensation of instantons
is responsible for the axion mass.

Recently, some of us and collaborators have made a proposal that the JTA would be able to explain
the dual phenomenon to radiative corrections \cite{Gamboa:2008ne} and used this idea to compute the
fermionic determinant in the $QED_3$ case. This result was immediately extended to consider the use
of the JTA to study $QED_3$ with magnetic-like defects. By a careful treatment of the symmetries
of the system we suggested a geometrical interpretation for some debatable issues in the
Maxwell-Chern-Simons-monopole system, such as the induction of the non-conserved electric
current together with the Chern-Simons term, the deconfinement transition and the computation
of the fermionic determinant in the presence of Dirac string singularities \cite{Grigorio:2008gd}.
It is important to point that the main signature of the JTA is the rank-jumping of the
field tensor due to the defects condensation. However, this discontinuous change of the theory
still puzzles a few. It is another goal of this investigation to shed some light in this matter.

In the present work we hope to help clarify the above mentioned issues focusing in the analysis
of the structure of these two methods, i.e., JTA and ALBA, by working out an explicit example.
Introducing a new Generalized Poisson's Identity (GPI) for $p$-branes in arbitrary space-time
dimensions and the novel concept of Poisson-dual branes we show that in the specific limit where
the defects proliferate until they occupy the whole space these two approaches are dual equivalent
prescriptions to obtain an effective theory for the phase where the defects are completely
condensed, starting from the fundamental theory defined in the normal phase where the defects are
diluted.

\section{Setting the problem}

The example we will work here is the Maxwell theory in the presence of monopoles that eventually
condense, which serves as an abelian toy model that simulates quark confinement.

The Maxwell field $A_\mu$ minimally coupled to electric charges $e$ and non-minimally coupled
to magnetic monopoles $g$ is described by the following action:
\begin{equation}
S=S_0^M+S_{int}=-\int d^4x\,\frac{1}{4}(F_{\mu\nu}-F^M_{\mu\nu})^2-\int d^4x\,j_\mu A^\mu,
\label{eq:1}
\end{equation}
where $j^\mu=e\delta^\mu(x;L')$ ($\tilde{j}^\mu=g\delta^\mu(x;L)$) is the the electric (magnetic)
current, being $\delta^\mu(x;L')$ ($\delta^\mu(x;L)$) a $\delta$-distribution that localizes the
world line $L'$ ($L$) of the electric (magnetic) charge $e$ ($g$). $F^{\mu\nu}_M=g\tilde{\delta}^
{\mu\nu}(x;S):=\frac{g}{2}\epsilon^{\mu\nu\alpha\beta}\delta_{\alpha\beta}(x;S)$ is the
magnetic Dirac brane, with $\delta_{\mu\nu}(x;S)$ a $\delta$-distribution that localizes the world
surface $S$ of the Dirac string coupled to the monopole \cite{dirac} and has the current
$\tilde{j}_\mu$ in its border. The field $A_\mu$ experiences a jump of discontinuity as it crosses
$S$, hence $F_{\mu\nu}$ has a $\delta$-singularity over $S$ \cite{felsager} that exactly cancels
the one in $F^M_{\mu\nu}$ such that $F_{\mu\nu}-F^M_{\mu\nu}:=F^{obs}_{\mu\nu}$ is the regular
combination which expresses the observable fields $\v{E}$ and $\v{B}$. As we shall see, the quantum
field theory associated to this action has two different kinds of local symmetries: the first
one is the usual electromagnetic gauge symmetry, $A_\mu(x)\rightarrow A'_\mu(x)=A_\mu(x)+
\partial_\mu \Lambda(x)$, with integrable $\Lambda$, i.e., $[\partial_\mu,\partial_\nu]\Lambda=0$.
The second one corresponds to the freedom of moving the unphysical surface $S$ over the space:
\begin{align}
&F_{\mu\nu}^M\rightarrow F_{\mu\nu}^M\,'=F_{\mu\nu}^M+\partial_\mu\Lambda^M_\nu-
\partial_\nu\Lambda^M_\mu,\nonumber \\
&A_\mu\rightarrow A'_\mu=A_\mu+\Lambda^M_\mu,
\label{eq:2}
\end{align}
where $\Lambda^M_\mu=g\tilde{\delta}_\mu(x;V):=\frac{g}{3!}\epsilon_{\mu\nu\alpha\beta}
\delta^{\nu\alpha\beta}(x;V)$, being $\delta_{\mu\nu\rho}(x;V)$ a $\delta$-distribution that
localizes the volume $V$ spanned by the deformation $S\rightarrow S'$ (the boundary $\partial S$ of
$S$ is physical and is kept fixed in the transformation such that $\partial S=\partial S'$). We
name here this second kind of local symmetry as brane symmetry. Taking into account the current
conservation we see that the action (\ref{eq:1}) is invariant under gauge transformations. But
(\ref{eq:1}) changes under brane transformations as $\Delta S=S'_{int}-S_{int}=-\int d^4x\,j_\mu
\Lambda_M^\mu=-egn,\;n\in\mathbb{Z}$ so that (\ref{eq:2}) is not a symmetry of (\ref{eq:1}). But
being the Dirac string unphysical we should not be able to detect it experimentally. So we need to
impose some consistency condition to make the Dirac string physically undetectable within the
present formalism. We can do it only by means of a quantum argument: the phase factor appearing
in the partition function associated to (\ref{eq:1}) changes under brane symmetry as
$e^{iS}\rightarrow e^{iS\,'}=e^{i(S+\Delta S)}=e^{iS}e^{-iegn},\;n\in\mathbb{Z}$. It should be
clear now that to keep the physics unchanged under brane transformations the consistency condition
needed to impose is $e^{-iegn}\equiv 1,\;n\in\mathbb{Z}\Rightarrow eg\equiv 2\pi N,\;N\in
\mathbb{Z}$, which is the famous Dirac quantization condition \cite{dirac}, a possible explanation
for the charge quantization.

Now in order to consider the monopole condensation (which will induce the electric charge
confinement) it is best to go to the dual picture. To obtain the dual action to (\ref{eq:1})
we introduce an auxiliary field $f_{\mu\nu}$ and define the master action by lowering the
order of the derivatives appearing in (\ref{eq:1}) via Legendre transformation:
\begin{equation}
\tilde{S}:=\int d^4x\left[-\frac{1}{2}f_{\mu\nu}F^{\mu\nu}_{obs}+\frac{1}{4}f_{\mu\nu}^2
-j_\mu A^\mu\right].
\label{eq:3}
\end{equation}
Extremizing $\tilde{S}$ with respect to $f_{\mu\nu}$ we get $f_{\mu\nu}=F_{\mu\nu}^{obs}$ and
substituting that in (\ref{eq:3}) we reobtain the original action (\ref{eq:1}) while
extremizing $\tilde{S}$ with respect to $A_\mu$ we get the condition $\partial_\mu f^{\mu\nu}=
j^\nu$, which can be solved by $f_{\mu\nu}\equiv\frac{1}{2}\epsilon_{\mu\nu\alpha\beta}
\tilde{F}_{obs}^{\alpha\beta}:=\frac{1}{2}\epsilon_{\mu\nu\alpha\beta}(\tilde{F}^{\alpha\beta}
-\tilde{F}_E^{\alpha\beta})$. We introduced the dual vector potential $\tilde{A}_\mu$ in
$\tilde{F}_{\mu\nu}:=\partial_\mu\tilde{A}_\nu-\partial_\nu\tilde{A}_\mu$ and the electric
Dirac brane $\tilde{F}^E_{\mu\nu}$ that localizes the world surface of the electric Dirac
string coupled to the electric charge. Substituting this result in (\ref{eq:3}) and discarding an
electric brane-magnetic brane contact term that does not contribute to the partition function due
to the Dirac quantization condition, we obtain the dual action:
\begin{equation}
\tilde{S}=\tilde{S}_0^E+\tilde{S}_{int}=\int d^4x\left[-\frac{1}{4}\tilde{F}_{\mu\nu}^{obs}\,^2-
\tilde{A}^\mu\tilde{j}_\mu\right],
\label{eq:4}
\end{equation}
where the couplings are inverted relatively to the ones in the original action (\ref{eq:1}): here
the dual vector potential $\tilde{A}_\mu$ couples minimally with the monopole and non-minimally
with the electric charge.

\section{Abelian Lattice Based Approach}

We are now in position to consider monopole condensation by applying the ALBA to the dual Maxwell
action (\ref{eq:4}). The main goal of this approach is to obtain an effective action for the
condensed phase in the dual picture. The ALBA is based on the observation that upon condensation,
the magnetic defects initially described by $\delta$-distributions are elevated to the field
category describing the long-wavelength fluctuations of the magnetic condensate. The condition
triggering the complete condensation of the defects is given by the disappearance of the
Poisson-dual brane (defined below) coming from a Generalized Poisson's Identity (see the discussion
in Appendix \ref{sec:app}).

We suppose that for the electric charges there are only a few fixed (external) worldlines $L'$
while for the monopoles we suppose that there is a fluctuating ensemble of closed worldlines $L$
that can eventually proliferate (the details of how such a proliferation takes place is a dynamical
issue not addressed neither by the ALBA nor by the JTA). The magnetic current is written in terms
of the magnetic Dirac brane as $\tilde{j}^\sigma=\frac{1}{2}\epsilon^{\sigma\rho\mu\nu}
\partial_\rho F^M_{\mu\nu}$. In order to allow the monopoles to proliferate we must give dynamics
to their magnetic Dirac branes since the proliferation of them is directly related to the
proliferation of the monopoles and their worldlines. Thus we supplement the dual action (\ref{eq:4})
with a kinetic term for the magnetic Dirac branes of the form $-\frac{\epsilon_c}{2}
\tilde{j}_\mu^2$, which preserves the local gauge and brane symmetries of the system. This
is an activation term for the magnetic loops. Hence, the complete partition function associated to
the extended dual action reads:
\begin{align}
&Z^c:=\int\mathcal{D}\tilde{A}_\mu\,\delta[\partial_\mu\tilde{A}^\mu]
e^{i\tilde{S}_0^E}Z^c[\tilde{A}_\mu],
\label{eq:5}
\end{align}
where the Lorentz gauge has been adopted for the dual gauge field $\tilde{A}_\mu$ and the partition
function for the brane sector $Z^c[\tilde{A}_\mu]$ is given by,
\begin{align}
&Z^c[\tilde{A}_\mu]:=\sum_{\left\{L\right\}}\delta[\partial_\mu\tilde{j}^\mu]
\exp\left\{i\int d^4x\left[-\frac{\epsilon_c}{2}\tilde{j}_\mu^2+\tilde{j}_\mu\tilde{A}^\mu\right]
\right\}.
\label{eq:6}
\end{align}
where the functional $\delta$-distribution enforces the closeness of the monopole worldlines.

Next, use is made of the Generalized Poisson's Identity (GPI) \cite{tese-marcelo} (see eq.
(\ref{eq:a6}) in Appendix \ref{sec:app}) in $d=4$
\begin{equation}
\sum_{\left\{L\right\}}\delta[\eta_\mu(x)- \delta_\mu(x;L)]=
\sum_{\left\{\tilde{V}\right\}}e^{2\pi i\int d^4x\, \tilde{\delta}_\mu(x;\tilde{V})
\eta^\mu(x)},
\label{eq:7}
\end{equation}
where $L$ is a 1-brane and $\tilde{V}$ is the 3-brane of complementary dimension. The GPI works as
an analogue of the Fourier transform: when the lines $L$ in the left-hand side of (\ref{eq:7})
proliferate, the volumes $\tilde{V}$ in the right hand side become diluted and vice versa (see the
discussion in Appendix \ref{sec:app}). We shall say that the branes $L$ and $\tilde{V}$ (or the
associated currents $\delta_\mu(x;L)$ or $\tilde{\delta}_\mu(x;\tilde{V})$) are Poisson-dual to
each other. Using (\ref{eq:7}) we can rewrite (\ref{eq:6}) as:
\begin{align}
Z^c[\tilde{A}_\mu]&=\int\mathcal{D}\eta_\mu\,\sum_{\left\{L\right\}}
\delta\left[g\left(\frac{\eta_\mu}{g}-\delta_\mu(x;L)\right)\right]\times\nonumber\\
&\times\delta\left[g\left(\partial_\mu\frac{\eta_\mu}{g}\right)\right]\exp\left\{
i\int d^4x\left[-\frac{\epsilon_c}{2}\eta_\mu^2+\eta_\mu\tilde{A}^\mu\right]\right\}\nonumber\\
&=\int\mathcal{D}\eta_\mu\,\sum_{\left\{\tilde{V}\right\}}
e^{2\pi i\int d^4x\,\tilde{\delta}_\mu(x;\tilde{V})\frac{\eta^\mu}{g}}\int
\mathcal{D}\tilde{\theta}\,\times\nonumber\\
&\times e^{i\int d^4x\,\tilde{\theta}\partial_\mu\frac{\eta^\mu}{g}}
\exp\left\{i\int d^4x\left[-\frac{\epsilon_c}{2}\eta_\mu^2+\eta_\mu\tilde{A}^\mu\right]\right\}
\nonumber\\
&=\sum_{\left\{\tilde{V}\right\}}\int\mathcal{D}
\tilde{\theta}\,\int\mathcal{D}\eta_\mu\,\exp\left\{i\int d^4x\left[
-\frac{\epsilon_c}{2}\eta_\mu^2+\right.\right.\nonumber\\
&\left.\left.-\eta^\mu\frac{1}{g}
(\partial_\mu\tilde{\theta}-\tilde{\theta}_\mu^V-g\tilde{A}_\mu)\right]\right\}.
\label{eq:8}
\end{align}
In the first line we introduced the auxiliary field $\eta_\mu$ which will replace the
$\delta$-distribution current in the condensed phase as discussed above. In the second line we
exponentiated the current conservation condition through use of the $\tilde{\theta}$ field and
also made use of the GPI to bring into the game the Poisson-dual current $\tilde{\theta}_
\mu^V=2\pi\tilde{\delta}_\mu(x;\tilde{V})$. We also made an integration by parts and discarded a
constant multiplicative factor since it drops out in the calculation of correlation functions.

Integrating the auxiliary field $\eta_\mu$ in the partial partition function (\ref{eq:8}) and
substituting the result back in the complete partition function (\ref{eq:5}) we obtain, as the
effective total action for the condensed phase in the dual picture, the London limit of the Dual
Abelian Higgs Model (DAHM):
\begin{equation}
\tilde{S}^L_{DAHM}=\int d^4x\left[-\frac{1}{4}\tilde{F}_{\mu\nu}^{obs}
\,^2+\frac{m_{\tilde{A}}^2}{2g^2}(\partial_\mu\tilde{\theta}-\tilde{\theta}_\mu^V
-g\tilde{A}_\mu)^2\right],
\label{eq:9}
\end{equation}
where we defined $m_{\tilde{A}}^2:=\frac{1}{\epsilon_c}$. This effective action is the main
result of this approach. In the next section we shall dualize this result and one could be concerned
with the fact that (\ref{eq:9}) constitutes a nonrenormalizable theory, thus requiring a cutoff
in order to be well defined as an effective quantum theory. However, one can always think of its
UV completion, in this case the complete DAHM, which is renormalizable, and then take its dual,
taking the London limit afterwards \cite{ripka}. At least in the case considered here, the result
is exactly the same one obtains by directly dualizing the London limit (\ref{eq:9}) of the DAHM,
thus justifying the procedure we shall adopt in the next section.

Considering now that a complete condensation of monopoles takes place we let their worldlines $L$
proliferate and occupy the whole space, implying that $\tilde{\theta}_\mu^V\rightarrow 0$ as seen
from (\ref{eq:7}) and the discussion afterwards (notice that $\tilde{\theta}_\mu^V$ appears as a
vortex-like defect for the scalar field $\tilde{\theta}$ describing the magnetic condensate, being
a parameter that controls the monopole condensation). Integrating the Higgs field $\tilde{\theta}$
we get a transverse mass term for $\tilde{A}_\mu$ (Higgs Mechanism) such that the electric field
has a finite penetration depth $\lambda=\frac{1}{m_{\tilde{A}}}=\sqrt{\epsilon_c}$ in the DSC: this
is the dual Meissner effect. Integrating now the field $\tilde{A}_\mu$ we obtain after some algebra
the effective action:
\begin{equation}
\tilde{S}_{eff}=\int d^4x\left[-\frac{m_{\tilde{A}}^2}{4}\tilde{F}_{\mu\nu}^E
\frac{1}{\partial^2+m_{\tilde{A}}^2}\tilde{F}^{\mu\nu}_E-\frac{1}{2}j_\mu
\frac{1}{\partial^2+m_{\tilde{A}}^2}j^\mu\right].
\label{eq:10}
\end{equation}

The first term in (\ref{eq:10}) is responsible for the charge confinement: it spontaneously breaks
the electric brane symmetry such that the electric Dirac string $\tilde{F}_{\mu\nu}^E$ acquires
energy becoming physical and constitutes now the electric flux tube connecting two charges of
opposite sign immersed in the DSC. The flux tube has a thickness equal to the penetration depth of
the electric field in the DSC: $\lambda=\frac{1}{m_{\tilde{A}}}=\sqrt{\epsilon_c}$. The shape of the
Dirac string is no longer irrelevant: the stable configuration that minimizes the energy is that
of a straight tube (minimal space). Substituting in the first term of (\ref{eq:10}) such a solution
for the string term, $\tilde{F}_{\mu\nu}^E=\frac{1}{2}\epsilon_{\mu\nu\alpha\beta}\frac{1}
{n\cdot\partial}(n^\alpha j^\beta-n^\beta j^\alpha)$, where $n^\mu:=(0,\v R:=\v R_1-\v
R_2)$ is a straight line connecting $+e$ in $\v R_1$ and $-e$ in $\v R_2$, and taking the static
limit we obtain a linear confining potential between the electric charges \cite{ripka}. We also
note that eliminating the magnetic condensate (i.e., taking the limit $m_{\tilde{A}}\rightarrow 0$)
we recover the diluted phase with no confinement: the interaction between the electric currents in
(\ref{eq:10}) becomes of the long-range (Coulomb) type and the confining term goes to zero (in
terms of the flux tube we see that it acquires an infinite thickness such that the electric field
is no longer confined and occupies the whole space).

In summary, the supplementing of the dual action with a kinetic term for the magnetic Dirac branes
which respects the local symmetries of the system, the subsequent use of the GPI (\ref{eq:a6}) and
the consideration of the limit where the Poisson-dual current $\tilde{\theta}_\mu^V$ goes
to zero gives us a proper condition for the complete condensation of monopoles, leading to
confinement, as viewed from the dual picture.

\section{Julia-Toulouse Approach}

Now we want to analyze the monopole condensation within the direct picture, where the defects
couple non-minimally with the gauge field $A_\mu$.

Using the Dirac quantization condition we can rewrite (\ref{eq:1}) as:
\begin{equation}
S=\int d^4x\left[-\frac{1}{4}F^{obs}_{\mu\nu}\,^2-\frac{1}{4}F_{obs}^{\mu\nu}
\epsilon_{\mu\nu\alpha\beta}\tilde{F}_E^{\alpha\beta}\right].
\label{eq:11}
\end{equation}

Julia and Toulouse made the crucial observation that if the monopoles completely condense we have
a complete proliferation of the magnetic strings associated to them, hence the field $A_\mu$ can
not be defined anywhere in the space. This implies that $F^{obs}_{\mu\nu}$ can no longer be written
in terms of $A_\mu$. The JTA consists in the rank-jump \textit{ansatz} of taking the object
$F^{obs}_{\mu\nu}$ as being the fundamental field describing the condensed phase. Hence $F^{obs}_
{\mu\nu}$ acquires a new meaning and becomes the field describing the magnetic condensate. Defining
$F^{obs}_{\mu\nu}:=-m_\Lambda\Lambda_{\mu\nu}$ and supplementing (\ref{eq:11}) with a kinetic term
of the form $\frac{1}{12}(\partial_\mu\Lambda_{\alpha\beta}+\partial_\beta\Lambda_{\mu\alpha}+
\partial_\alpha\Lambda_{\beta\mu})^2$ for the new field $\Lambda_{\mu\nu}$, we obtain as the
effective action for the condensed phase, in the direct picture, the massive Kalb-Ramond action:
\begin{equation}
S_{K-R}=\int d^4x\left[-\frac{1}{2}(\partial_\mu\tilde{\Lambda}^{\mu\nu})^2+
\frac{m_\Lambda^2}{4}\tilde{\Lambda}_{\mu\nu}^2+\frac{m_\Lambda}{2}\tilde{\Lambda}^{\mu\nu}
\tilde{F}^E_{\mu\nu}\right],
\label{eq:12}
\end{equation}
where $\tilde{\Lambda}^{\mu\nu}:=\frac{1}{2}\epsilon^{\mu\nu\alpha\beta}\Lambda_{\alpha\beta}$.

Notice that in implementing the JTA the fundamental field of the theory experiences a rank-jump
through the phase transition: we started with a 1-form in the normal phase and finished with a
2-form in the completely condensed phase. The rank-jump is a general feature of the JTA since in
implementing this prescription we always use the \textit{ansatz} of reinterpretating the kinetic
term with non-minimal coupling for the field describing the diluted phase as being a mass term
for the new field describing the condensate formed in the phase where the defects proliferate until
occupy the whole space.

Let us now apply the duality transformation in (\ref{eq:9}). For this we introduce an auxiliary
field $f_{\mu\nu}$ such that the master action reads:
\begin{align}
S_{Master}&:=\int d^4x\left[-\frac{1}{2}f_{\mu\nu}(\tilde{F}^{\mu\nu}-\tilde{F}_E^{\mu\nu})+
\frac{1}{4}f_{\mu\nu}^2+\right.\nonumber\\
&\left.+\frac{m_{\tilde{A}}^2}{2g^2}(\partial_\mu\tilde{\theta}-\tilde{\theta}_\mu^V-g\tilde{A}_\mu)
^2\right].
\label{eq:13}
\end{align}

Extremizing (\ref{eq:13}) with respect to $f_{\mu\nu}$ we get $f_{\mu\nu}=\tilde{F}_{\mu\nu}^{obs}$
and substituting this result back in the master action we recover (\ref{eq:9}). On the other hand,
extremizing (\ref{eq:13}) with respect to $\tilde{A}_\mu$ we obtain:
\begin{equation}
\tilde{A}^\nu=-\frac{1}{m_{\tilde{A}}^2}\partial_\mu f^{\mu\nu}+
\frac{1}{g}(\partial^\nu\tilde{\theta}-\tilde{\theta}^\nu_V).
\label{eq:14}
\end{equation}

Substituting (\ref{eq:14}) in (\ref{eq:13}), it follows that:
\begin{align}
S_{Master}&=\int d^4x\left[-\frac{1}{2m_{\tilde{A}}^2}(\partial_\mu f^{\mu\nu})^2+
\frac{1}{4}f_{\mu\nu}^2+\frac{1}{2}f_{\mu\nu}\tilde{F}_E^{\mu\nu}+\right.\nonumber\\
&\left.+\frac{1}{g}\partial_\mu\tilde{\theta}_\nu^V f^{\mu\nu}\right],
\label{eq:15}
\end{align}
where we integrated by parts and considered the antisymmetry of $f^{\mu\nu}$ in order to use
$\partial_\mu\partial_\nu f^{\mu\nu}=0$.

Defining now $f_{\mu\nu}:=m_{\tilde{A}}\tilde{\Lambda}_{\mu\nu}$ and making the identification
$m_{\tilde{A}}\equiv m_{\Lambda}$, we get as the dual action to (\ref{eq:9}) the massive
Kalb-Ramond action in the presence of vortices, a generalization of the result obtained by Quevedo
and Trugenberger in \cite{QT}:
\begin{align}
S_{KR}^V&=\int d^4x\left[-\frac{1}{2}(\partial_\mu \tilde{\Lambda}^{\mu\nu})^2+
\frac{m_\Lambda^2}{4}\tilde{\Lambda}_{\mu\nu}^2+\frac{m_\Lambda}{2}
\tilde{\Lambda}_{\mu\nu}\tilde{F}_E^{\mu\nu}+\right.\nonumber\\
&\left.+\frac{m_\Lambda}{2g}(\partial_\mu\tilde{\theta}_\nu^V-
\partial_\nu\tilde{\theta}_\mu^V)\tilde{\Lambda}^{\mu\nu}\right].
\label{eq:16}
\end{align}

More precisely, this extension consists in the construction of an action for the case with an
incomplete condensate that is however already described by a rank-jumped tensor. If we now take
the limit $\tilde{\theta}_\mu^V\rightarrow 0$ in (\ref{eq:16}) we recover exactly the massive
Kalb-Ramond action (\ref{eq:12}) obtained in \cite{QT} through the application of the JTA to
(\ref{eq:1}). That establishes the duality between the JTA and the ALBA in the limit where the
Poisson-dual current goes to zero, which physically corresponds to the limit of complete
condensation of the defects. However, (\ref{eq:16}) with $\tilde{\theta}_\mu^V\neq 0$ displays
a new and important result, which is a consequence of this formalism, showing that the rank-jump
which is the signature of the JTA also occurs in the partial condensation process with the presence
of vortex-like defects.

\section{Conclusion}

We established the equivalence through duality of two different approaches developed to handle
defects, represented by magnetic monopoles in the example worked here, in the physically
interesting context where the defects dominate the dynamics of the system. It was clearly shown
that the two approaches are complementary, being different descriptions of the same phenomenon in
the limit where the Poisson-dual current vanishes which characterizes the complete
condensation of the defects. Indeed, within the formalism here called as ALBA the transition
becomes smoother since the Poisson-dual current $\tilde{\theta}_\mu^V$ appears as a
parameter that controls the proliferation of the magnetic defects. On the other hand, within the
formalism referred to as JTA the phase transition is signalized by a rank-jump of the tensor field
and seems to be a discontinuous phenomenon. However, the duality JTA-ALBA brings a new possibility.

It is important to say that this dual equivalence was possible due to a suitable interpretation of
the generalization of the Poisson identity developed here. We clearly showed that this identity is
an essential tool to use in the context of defects condensation: the proliferation of the branes in
one of the sides of the identity is accompanied by the dilution of the branes of complementary
dimension in the other side of the identity. Due to this observation we were able to identify the
signature of the complete condensation of defects in the dual picture (ALBA) with the vanishing of
the Poisson-dual current. As the main result, we showed that in this specific limit, when
the Poisson-dual current is set to zero, the ALBA and the JTA are two dual equivalent
prescriptions for describing condensation of defects.

As the final remark we point out the fact that when we consider nonzero configurations of the
Poisson-dual current $\tilde{\theta}_\mu^V$ we allow the description of an intermediary
region interpolating between the diluted and the completely condensed phases. As discussed, this
corresponds to the presence of vortex-like defects in the condensate. It is possible to see that
this new phase with the presence of vortices ($\tilde{\theta}_\mu^V\neq 0$), just like in the
extreme case where the complete monopole condensation sets in, is also described within the direct
picture by a rank-jumped action. The JTA as originally described by Quevedo and Trugenberger,
therefore, will describe the physically interesting extreme case where all defects are condensed.

\section{Acknowledgements}
We thank Conselho Nacional de Desenvolvimento Cient\'ifico e Tecnol\'ogico (CNPq),
Coordena\c{c}\~ao de Aperfei\c{c}oamento de Pessoal de N\'ivel Superior (CAPES) and Funda\c{c}\~ao
de Amparo \`a Pesquisa do Estado do Rio de Janeiro (FAPERJ) for financial support.

\appendix
\section{Generalized Poisson's Identity (GPI)}
\label{sec:app}
In this appendix we generalize the reasoning used in \cite{mvf} in order to account for an
ensemble of $p$-branes in arbitrary space-time dimensions.

Let us consider a $d$-dimensional hypercubical lattice with spacing $a$. Attribute to each site
$x=(x_1,x_2,\ldots,x_d),\,x_1,x_2,\ldots,x_d\in a\mathbb{Z}$ of the lattice a configuration
\begin{equation}
\vartheta_i^V(x):=2\pi\frac{n_i(x)}{a^p},
\label{eq:a1}
\end{equation}
where $p\le d,\,p,d\in\mathbb{N}$ and $i$ is a set of $k\le d,\,k\in\mathbb{N}$ indices each
one of them running from 1 to $d$ and $n_i(x)\in\mathbb{Z}$.

The Poisson's summation formula is given by
\begin{equation}
\sum_{n\in\mathbb{Z}}e^{2\pi inf}=\sum_{m\in\mathbb{Z}}\delta(f-m),
\label{eq:a2}
\end{equation}
where $f$ is a integrable function.

Using (\ref{eq:a2}) for each pair $(x,i)$ it follows that
\begin{align}
&\sum_{\left\{n_i(x)\in\mathbb{Z}\right\}}\exp\left[2\pi i\sum_x a^d\frac{n_i(x)}{a^p}f_i(x)\right]=
\nonumber \\
&=\sum_{\left\{m_i(x)\in\mathbb{Z}\right\}}\prod_{(x,i)}\delta\left(f_i(x)-\frac{m_i(x)}{a^{d-p}}
\right),
\label{eq:a3}
\end{align}
where we have used the fact that the exponential argument must be nondimensional, hence
$a^{d-p+\left[f\right]}\equiv a^0=1\Rightarrow\left[f\right]=a^{p-d}$.

The continuum limit corresponds to make the number $N$ of lattice sites go to infinity while
keeping the lattice hypervolume $V_d$ fixed which gives the condition $a\rightarrow 0$. In this
limit we formally define the Poisson-dual current by
\begin{equation}
\theta_i^V(x;\xi^p):=\lim_{\substack{a\rightarrow 0 \\ N\rightarrow\infty \\ V_d\,\mbox{cte}}}
\vartheta_i^V(x)=2\pi\lim_{\substack{a\rightarrow 0 \\ N\rightarrow\infty \\ V_d\,\mbox{cte}}}
\frac{n_i(x)}{a^p}.
\label{eq:a4}
\end{equation}

The object $\theta_i^V(x;\xi^p)$ has dimension $a^{-p}$ and is singular over a region $\xi^p$ of
dimension $p$ on the lattice where $\{n_i(x\in\xi^p):\neq 0\}$. In the rest of the lattice, where
$\{n_i(x\notin\xi^p):=0\}$, we have from (\ref{eq:a1}) that $\vartheta_i^V(x)=0$ such that
$\theta_i^V(x;\xi^p)$ vanishes outside the region $\xi^p$. Thus we identify the object
$\theta_i^V(x;\xi^p)$ with a delta configuration that localizes the $p$-brane $\xi^p$:
\begin{equation}
\theta_i^V(x;\xi^p)=2\pi\delta_i(x;\xi^p).
\label{eq:a5}
\end{equation}

Hence in the continuum limit the identity (\ref{eq:a3}) is given by
\begin{equation}
\sum_{\left\{\xi^p\right\}}e^{2\pi i\int d^dx\delta_i(x;\xi^p)f_i(x)}=
\sum_{\left\{\chi^{d-p}\right\}}\delta\left[f_i(x)-\delta_i(x;\chi^{d-p})\right],
\label{eq:a6}
\end{equation}
which is the GPI.

The brane proliferation-dilution interpretation of the GPI (\ref{eq:a6}) follows from the following
reasoning: if $\left\{\chi^{d-p}\right\}\rightarrow\emptyset$ then $\delta_i(x;\chi^{d-p})
\rightarrow 0$ (there are no $\{\chi^{d-p}\}$ branes in the space to be localized) and
\begin{equation}
\sum_{\left\{\chi^{d-p}\right\}}\delta\left[f_i(x)-\delta_i(x;\chi^{d-p})\right]\rightarrow\delta
\left[f_i\right]\equiv\int\mathcal{D}\tau_i e^{i\int d^dx\tau_i f_i}.
\label{eq:a7}
\end{equation}

Comparing (\ref{eq:a6}) and (\ref{eq:a7}) we see that in the limit of dilution of the $\{\chi^{d-p}
\}$ branes we have $\theta_i^V(x;\xi^p)=2\pi\delta_i(x;\xi^p)\rightarrow\tau_i$ and $\sum_{
\left\{\xi^p\right\}}\rightarrow\int\mathcal{D}\tau_i$ which corresponds to the proliferation of
the $\{\xi^p\}$ branes.

Conversely, in the limit of proliferation of the $\{\chi^{d-p}\}$ branes, $\theta_i^V(x;\chi^{d-p})=
2\pi\delta_i(x;\chi^{d-p})\rightarrow\gamma_i$ and $\sum_{\left\{\chi^{d-p}\right\}}\rightarrow
\int\mathcal{D}\gamma_i$ we have
\begin{equation}
\sum_{\left\{\chi^{d-p}\right\}}\delta\left[f_i(x)-\delta_i(x;\chi^{d-p})\right]\rightarrow\int
\mathcal{D}\gamma_i\delta\left[f_i-\gamma_i\right]=\mathds{1}.
\label{eq:a8}
\end{equation}

Comparing (\ref{eq:a6}) and (\ref{eq:a8}) we see that in the limit of proliferation of the
$\{\chi^{d-p}\}$ branes we have $\theta_i^V(x;\xi^p)=2\pi\delta_i(x;\xi^p)\rightarrow 0$ which
corresponds to the dilution of the $\{\xi^p\}$ branes.

It is important to notice that the information about which brane configurations are accessible by
the system in the brane sums in the GPI (\ref{eq:a6}) is not present in this formulation, being an
external input controlled by hands as when we considered previously, for example, the extreme cases
of prolific or diluted accessible brane configurations.

\end{document}